\documentclass[12pt]{article}
\pdfoutput=1
\usepackage{amsmath,amssymb,amsfonts,epsfig,graphicx,euscript}%
\usepackage[pdftex,bookmarks,bookmarksnumbered,linktocpage,pdfstartview=FitH]{hyperref}
\hypersetup{colorlinks,%
citecolor=red,%
filecolor=blue,%
linkcolor=blue,%
urlcolor=blue,%
pdftex}
\usepackage[all]{hypcap}
\numberwithin{equation}{section}
\usepackage{cite}
\newcommand{\bse}{\begin{subequations}}
\newcommand{\ese}{\end{subequations}}
\newcommand{\be}{\begin{equation}}
\newcommand{\ee}{\end{equation}}
\newcommand{\bea}{\begin{eqnarray}}
\newcommand{\eea}{\end{eqnarray}}
\newcommand{\ba}{\begin{array}}
\newcommand{\ea}{\end{array}}

\newcommand{\A}{(2 \pi \alpha')}
\newcommand{\s}{\sigma}

\begin{document}
\hfill%
\vbox{
    \halign{#\hfil        \cr
           IPM/P-2012/046\cr
                     }
      }
\vspace{1cm}
\begin{center}
{ \Large{\textbf{Thermalization
 in External Magnetic Field}}} 
\vspace*{1.8cm}
\begin{center}
{\bf Mohammad Ali-Akbari\footnote{aliakbari@theory.ipm.ac.ir}, Hajar Ebrahim\footnote{hebrahim@ipm.ir}}\\%
\vspace*{0.4cm}
{\it {${}^1$School of Particles and Accelerators,\\ ${}^2$School of Physics,\\ Institute for Research in Fundamental Sciences (IPM),\\
P.O.Box 19395-5531, Tehran, Iran}}  \\

\vspace*{1.5cm}
\end{center}
\end{center}

\vspace{.5cm}
\bigskip
\begin{center}
\textbf{Abstract}
\end{center}
In the AdS/CFT framework meson thermalization in the presence of a constant external magnetic field in a strongly coupled gauge theory has been studied. In the gravitational description the thermalization of mesons corresponds to the horizon formation on the flavour D7-brane which is embedded in the $AdS_5\times S^5$ background in the probe limit.  The apparent horizon forms due to the time-dependent change in the baryon number chemical potential, the injection of baryons in the gauge theory. We will numerically show that the thermalization happens even faster in the presence of the magnetic field on the probe brane. We observe that this reduction in the thermalization time sustains up to a specific value of the magnetic field.

\newpage

\tableofcontents

\section{Introduction and Results}
Quark-gluon plasma
created by colliding relativistic heavy ions such as Gold at
Relativistic Heavy Ion Collider or Lead at Large Hadron Collider, is
believed to be strongly coupled \cite{Shuryak:2003xe,Shuryak:2004cy}. In the initial
stages right after the collision the quark-gluon plasma is far from
equilibrium and hence it would be difficult to study its properties. But
after a very short time, ($\tau\leq1~{\rm{fm/c}}$), the hydrodynamics is
applicable indicating that the local equilibrium has been partly reached
\cite{Heinz:2004pj,Luzum:2008cw}. An important question is how a strongly coupled system
can rapidly relax from far from equilibrium regime to the
hydrodynamic phase. This process is generally called
rapid thermalization. Since the system is strongly coupled the usual
perturbative methods seem inapplicable to study such a process. Recent calculations show that AdS/CFT correspondence \cite{Maldacena} which loosely speaking is a duality between a strongly coupled system and a classical gravity theory can be a suitable framework to
describe rapid thermalization \cite{Chesler,Bhattacharyya,Heller:2011ju}.

An interesting discussion has been made in \cite{Kharzeev} indicating that after the collision of two heavy ions a strong magnetic field
is produced which is present only for a very short time. Since this magnetic field is nonzero in the initial stages of the collision it is reasonable to
study the process of thermalization in its presence. Therefore one can use the gauge/gravity duality to explain how the magnetic field can effect the rapid thermalization in QGP. This is exactly the question that we are trying to address in this paper.

According to the AdS/CFT correspondence the four-dimensional
superconformal Yang-Milles (SYM) theory with gauge group $SU(N_c)$
is dual to the type IIB string theory on $AdS_5\times S^5$ geometry which
describes the near horizon geometry of a stack of $N_c$ extremal
D3-branes \cite{Maldacena}. This correspondence is more understood in the large
$N_c$ and large t'Hooft coupling, $\lambda=g_{YM}^2 N_c$, limit. In
these limits the type IIb supergravity on $AdS_5\times S^5$ geometry
is dual to the strongly coupled SYM theory and hence it provides a
useful tool to study the strongly coupled regime of the SYM theory.
This correspondence is extended in various ways. In particular the
thermal SYM theory corresponds to the supergravity in the
AdS-Schwarzschild black hole background. The Hawking temperature of the black
hole is identified with the temperature of the thermal gauge theory
\cite{Maldacena}.

In the context of AdS/CFT correspondence matter fields (or quarks) in the fundamental representation are introduced in the gauge theory by embedding flavour D7-branes in the probe limit \cite{Karch}. By probe limit one means that the number of flavour D7-branes are much smaller than the
number of D3-branes. In this system open strings
stretched between D3-branes and D7-branes are realized as quarks in the SYM theory. Their mass is proportional
to the asymptotic separation of the D3 and D7-branes in transverse
directions. Additionally the open strings with both
endpoints on the D7-branes are considered as mesons. This system is a notable candidate to describe QCD-like
theories \cite{Erdmenger1}.

The process of thermalization refers to an increase in the temperature of a system from an initial value which can be zero to a higher  value in a non-adiabatic way which takes the system out of equilibrium. This is done by the injection of energy into the gauge theory or equivalently, in the context of AdS/CFT, by the collapse of matter in $AdS_5\times S^5$ background. This injection of energy can be done by turning on a time-dependent source term in the gauge theory. It has been shown that the addition of time-dependent sources such as metric or scalar field on the boundary result in horizon formation in the bulk \cite{Chesler,Bhattacharyya}.  These sources are coupled to the energy-momentum tensor and scalar operators, respectively. Therefore one can simply say that the thermalization in the field theory is dual to the black hole formation in the bulk.

So far we have been discussing the thermalization in the so-called gluon sector which corresponds to the horizon formation in the bulk. This can be generalized to the horizon formation on the probe branes which is usually referred to as thermalization in the meson sector \cite{Hashimoto,Das:2010yw,paper}. In \cite{Hashimoto} the energy injection that leads to the thermalization in gauge theory is modeled by the injection of baryons which corresponds to throwing the open strings on the probe D7-brane from the boundary. The time-scale of the thermalization is identified with the time-scale of the apparent horizon formation observed on the boundary. In such a set-up we turn on a constant magnetic field
on the D7-brane and study the meson
thermalization in the presence of this magnetic field.

The main result of this paper is that the thermalization time corresponding to the meson melting decreases as one increases the
external magnetic field on the probe D7-brane. This can be justified by an energy evaluation which will be explained in more details in the last section.
It has been observed that when the external magnetic field is zero the thermalization time scales as $(\frac{\lambda}{n_B^2 \omega^2})^{\frac{1}{8}}$
where $n_B$ is the baryon number density which is injected into the system and $\omega$ shows
the period of time that the injection happens \cite{Hashimoto} . In the presence of the magnetic field the difference between the thermalization times
of zero and non-zero magnetic fields behaves as%
\be %
 \delta t_{th}=t_{th}^B-t_{th}^{B=0}\thicksim\left(\lambda^7 n_B^2\omega^2\right)^{-\frac{1}{8}} B^\kappa,
\ee %
where for weak magnetic field, defined as $(2\pi\alpha') B\ll1$, $\kappa$ is 2 and as $B$ is raised it deviates from $2$. This model shows that  the presence of the magnetic field helps the thermalization, or in more exact words meson dissociation,  to happen even faster. We observe that this reduction in the thermalization time happens up to a specific value of the magnetic field which depends on the numeric values of the parameters of the theory.

\section{D7-Brane Embeddings in External Magnetic Field}\label{hajar1}
In AdS/CFT one can add the fundamental matter into the gauge theory by embedding a probe D7-brane in an asymptotically AdS background \cite{Karch}. The asymptotic behaviour of this brane teaches us about the fundamental matter (quark) mass and the condensation which indicates the spontaneous chiral symmetry breaking \cite{Kruczenski:2003uq}. One way to break the chiral symmetry is to embed the D7-brane in the $AdS_5\times S^5$ background with a nonzero constant Kalb-Ramond field\footnote{Note that a pure gauge Kalb-Ramond field which is a solution
to type IIb supergravity equations of motion can be replaced by the constant magnetic field $B$ on the D7-brane. This field content does not deform the $AdS_5\times S^5$ background \cite{Filev1}.} \cite{Erdmenger,Filev1,Filev2}. This has the effect of repelling the D7-brane from the origin. The nontrivial shape of the probe brane gives the condensation ($c$) which varies with respect to the quark mass.  $c$ with respect to $m$ shows a spiral behaviour near the origin. This means that for a given value of the quark mass a number of solutions exist. Among them the
physical solution is the most energetically favourable one. The presence of the constant Kalb-Ramond or equivalently constant magnetic field leads to nonzero quark condensate even at zero quark mass. In the
following we will discuss these results in more details.

The background of interest is $AdS_5\times S^5$ geometry describing the near horizon geometry of $N_c$ D3-branes
\be\begin{split}\label{metric}%
 ds^2&=(\frac{u}{R})^2(-dt^2+d\vec{x}^2)+(\frac{R}{u})^2(d\rho^2+\rho^2d\Omega_3^2+d\sigma^2+\sigma^2d\varphi^2), \cr
 g_sC_{(4)}&=\frac{u^4}{R^4}dx^0\wedge dx^1\wedge dx^2\wedge dx^3,\ g_s=e^{\phi_\infty},\ R^4=4\pi g_s N_c l_s^4~,
\end{split}\ee %
where $C_{(4)}$ and $\phi$ are four-form and Dilaton fields, respectively. In this coordinate $u^2=\rho^2+\sigma^2$ and $l_s=\sqrt{\alpha'}$
is the string length scale. In order to introduce the fundamental matter we have to add a space-filling flavour D7-brane in the probe limit to this background. The configuration of the D3-D7 branes is
\be %
\begin{array}{ccccccccccc}
                   & 0 & 1 & 2 & 3 & 4 & 5 & 6 & 7 & 8 & 9 \\
                  D3 & \times & \times & \times & \times &  &  &  &  &  &  \\
                  D7 & \times & \times & \times & \times & \times & \times & \times & \times &  &
\end{array}
\ee %
In the probe limit the low energy effective action for a D7-brane in an arbitrary
background is described by Dirac-Born-Infeld
(DBI) and Chern-Simons (CS) actions %
\be\begin{split}\label{action} %
 S &= S_{\rm{DBI}}+ S_{\rm{CS}}~,\cr
 S_{{\rm{DBI}}}&=-\mu_7\int d^{8}\xi\
 e^{-\phi}\sqrt{-\det(g_{ab}+B_{ab}+2\pi\alpha'F_{ab})}~,\cr
 S_{\rm{CS}} &=\mu_7\int P[\Sigma C_{(n)}e^B]e^{2\pi\alpha'F}~,
\end{split}\ee %
where induced metric $g_{ab}$ and induced Kalb-Ramond field $B_{ab}$ are given by
\be\begin{split} %
 g_{ab}=G_{MN}\partial_a X^M\partial_b X^N,\cr
 B_{ab}=B_{MN}\partial_a X^M\partial_b X^N.
\end{split}\ee %
$\xi^a$ are worldvolume coordinates and $\mu_7^{-1}=(2\pi)^7l_s^8g_s$ is the D7-brane tension.
$G_{MN}$ is the background metric introduced in \eqref{metric} and in our background $B_{MN}$ is zero.  $F_{ab}$ is the field strength of the gauge field living on the D7-brane. In the CS part, $C_{(n)}$ is a $n$-form field and $P[...]$ denotes the
pull-back of the bulk space-time tensors to the D7-brane worldvolume.

We work in the static gauge where the D7-brane is extended along $t,\vec{x},\rho,\Omega_3$ and consider the magnetic field
\be %
  B=F_{xy},
\ee %
where $B$ is constant. The presence of this external magnetic field will break supersymmetry on the probe brane. $\sigma$ as a function of $\rho$ describes the shape of the brane. We set all other fields to zero. The D7-barne action then becomes %
\bea\label{hajaraction}
S_{DBI}=-\mu_7{\rm{V}}_3{\rm{Vol}}(\Omega^3) \int dt d\rho\ \rho^3 \sqrt{1+(2 \pi \alpha')^2 R^4 u^{-4} B^2} \sqrt{1+\sigma'^2},
\eea
where $\sigma'=\frac{\partial \sigma}{\partial \rho}$. The equation of motion for $\sigma(\rho)$ can be easily found %
\be\label{EOM}\begin{split} %
 \partial_\rho\bigg(\frac{\rho^3\sigma'}{\sqrt{1+\sigma'^2}}\sqrt{1+\frac{(2\pi\alpha')^2 B^2 R^4}{(\rho^2+\sigma^2)^2}}\ \bigg)
 +\frac{2\rho^3(2\pi\alpha')^2 B^2 R^4}{(\rho^2+\sigma^2)^3}
 \frac{\sigma\sqrt{1+\sigma'^2} }{\sqrt{1+\frac{(2\pi\alpha')^2 B^2 R^4}{(\rho^2+\sigma^2)^2}}}=0.
\end{split}\ee %
In the asymptotic limit where $\rho\rightarrow\infty$, the above equation reduces to
\be %
 \partial_\rho\bigg(\frac{\rho^3\sigma'}{\sqrt{1+\sigma'^2}}\bigg)=0,
\ee %
and hence its solution approaches
\be %
 \sigma(\rho)=m+\frac{c}{\rho^2}+....
\ee %
According to the AdS/CFT correspondence  the
leading and sub-leading terms in the asymptotic expansion of various bulk fields
define the source for the dual operator and the expectation value of that operator, respectively. For $\sigma(\rho)$ the leading term corresponds
to the quark mass, $m_q=m/(2\pi\alpha')$, and the
sub-leading term is related to the chiral condensate via $c=(2\pi\alpha')^3\langle \bar{q}q\rangle$ \cite{Kruczenski:2003uq}.

\begin{figure}[ht]
\begin{center}
\includegraphics[width=2.6 in]{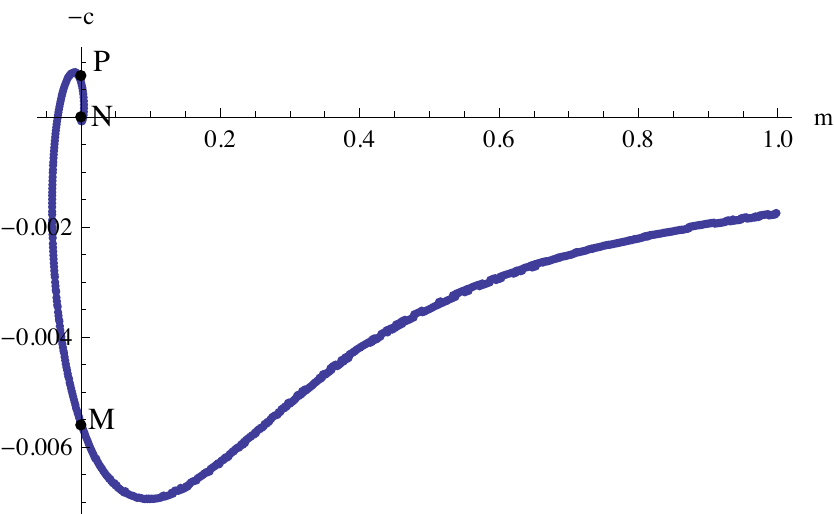}
\hspace{2mm}
\includegraphics[width=2.6 in]{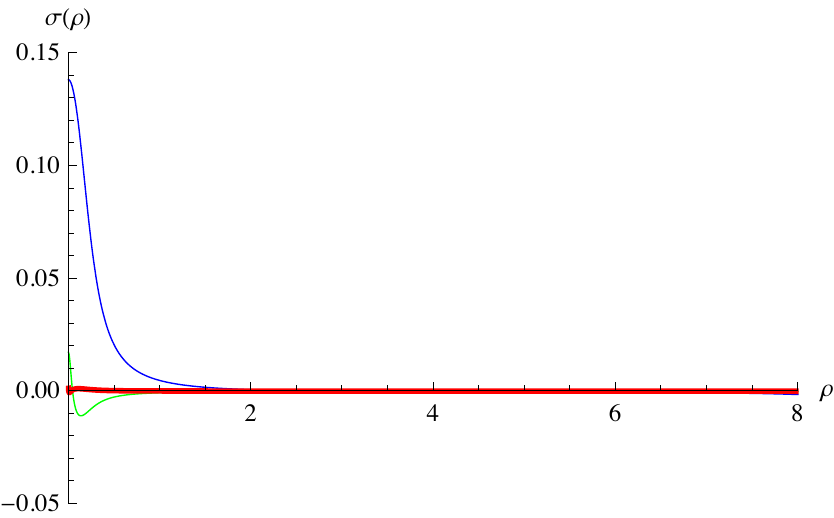}
\caption{Left: The condensation as a function of the mass for $2\pi\alpha'B=0.09$
and $R=1$. Points M, N and P represent three massless solutions. Right: The shape of three massless solutions corresponding to the points M, N and P.  The blue curve (representing point M) is the physical massless solution.
 \label{massless}}
\end{center}
\end{figure}

The equation of motion for $\sigma(\rho)$ is complicated to be solved
analytically and hence we use numerical methods to find its solutions.
In order to solve it numerically we have to impose two boundary conditions at a specific
point which we choose them to be
$\sigma(0)\geq0$ and $\sigma'(0)=0$. The second condition
guarantees the regularity of the D7-brane embedding. The above
boundary conditions for a \textit{fixed} value of the magnetic field and
different values of $\sigma(0)$ leads to different values of  mass and
condensation. The final results are plotted in figure
\ref{massless}(left). Evaluating the energy for different configurations of $c$ and $m$ shows that
the region with $m>0$ and $-c<0$, the bottom right quadrant in the
figure \ref{massless}(left), describes physical solutions. It can be seen that  in the presence of a fixed magnetic field
there are three massless solutions. One of them is  trivial
and  is zero for all values of $\rho$
indicating that $m=c=0$. Although for the other two solutions $m$ is still
zero, the value of $c$ is not zero. A nonzero $c$
means that chiral symmetry has been spontaneously broken by the external magnetic
field \cite{Erdmenger}. Among massless solutions in figure \ref{massless}(left) the solution
corresponding to the point $M$ is energetically more favourable. $\sigma(0)$ at this point is
$0.139$ where $2\pi\alpha'B=0.09$. The blue, red and green curves in figure \ref{massless}(right)
represent brane embeddings with asymptotic behaviours corresponding to the points $M$, $N$ and $P$, respectively.

\begin{figure}[ht]
\begin{center}
\includegraphics[width=2.9 in]{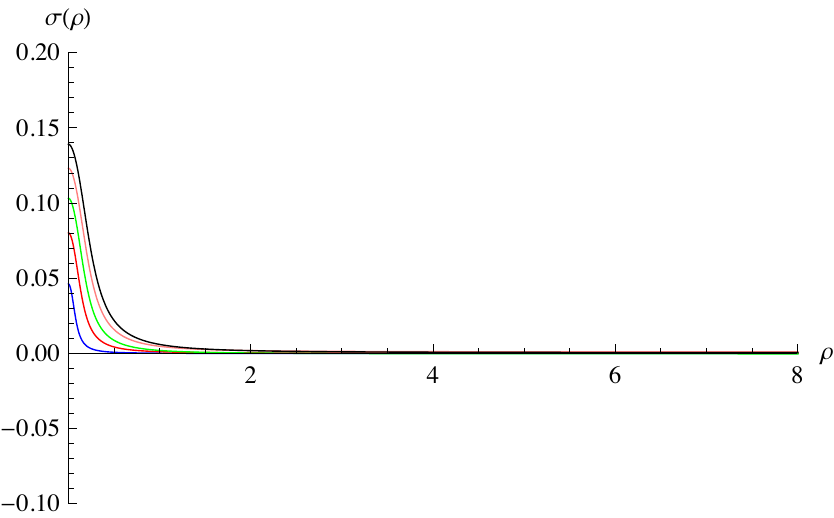}
\caption{Different physical massless solutions for
$(2\pi\alpha')B=0.09,0.07,0.05,0.03$ and $0.01$ (top to bottom) and
$R=1$. \label{weak}}
\end{center}
\end{figure}

In figure \ref{weak} physical massless configurations for various
values of the magnetic field are plotted. It is clear that
for larger values of the magnetic field  $\sigma(0)$
also becomes larger. In other words the turning point of the
D7-brane configurations pushes more and more towards the boundary as one
raises $B$.

An important observation is that for small values of the
magnetic field the deformation of the D7-brane is not significant.
Therefore $\sigma(\rho)\thickapprox0$ is an acceptable approximation
in the weak magnetic field limit. We will use this fact in the
subsection \ref{hajar} to simplify the calculation of  the thermalization
time.

\section{Meson Thermalization in External Magnetic Field}
The process in which the temperature of a system changes non-adiabatically from $T_1$ (which can be zero or non-zero) to
a larger value $T_2$ is generally called thermalization. In the context of
AdS/CFT a zero temperature state in a strongly coupled gauge theory
is dual to the pure $AdS_5\times S^5$ background in the bulk.
Similarly a thermal state is dual to the $AdS$-black hole background.
Then the thermalization in the gauge theory side is identified by
the horizon formation in the gravity side. A generalization has been done in
\cite{Hashimoto}  where meson thermalization is identified by the apparent horizon
formation on the flavour D7-brane. The thermalization happens due to the
injection of baryons into the gauge theory. In the gravity side
baryon injection is presented by throwing the fundamental strings on
the D7-brane. The open strings stretched between
D3- and D7-branes play the role of the fundamental quarks in
the gauge theory side. A time dependent non-adiabatic baryon injection
will suddenly increase the number of baryon charges
in a time dependent manner and consequently produces a time
dependent chemical potential. The time dependence of the chemical
potential or baryon number density can finally lead to the horizon formation on the
D7-brane.

In the AdS/CFT correspondence the static chemical potential, $\mu$,
is introduced via time component of the gauge filed as
\be %
 \mu=\int_0^\infty d\rho~ \partial_\rho A_t(\rho),
\ee %
in the gauge where the radial component of the gauge field is set to
zero. In order to obtain a time dependent chemical potential we
have to consider a time dependent gauge filed configuration. This can be done by introducing
source terms to the DBI action
\be %
\delta S=\mu_7{\rm{V}}_3{\rm{Vol}}(\Omega_3)\int dt d\rho(A_tj^t+A_\rho j^\rho),
\ee %
which describes the dynamics of open strings, or equivalently,
fundamental quarks injected on the brane from the boundary.
Notice that the source terms are now time dependent and their
form determines how the chemical potential
varies in time.

Since we consider the massless quarks it is convenient to work with light-cone coordinates which are defined as
\be %
x^\pm=t\pm z=t\mp \int d\rho\sqrt{H(1+\sigma')},
\ee %
where $H=(R/u)^4$. Note that the boundary is located at $z=0$ and
the Poincare horizon is $z\rightarrow \infty$ . Without loss of
generality we assume the currents are only functions of $x^-$. More
precisely we consider $j^\rho= -
\big(H(1+\sigma')\big)^{-\frac{1}{2}}j^t= g'(x^-)$. Therefore in the
presence of the magnetic field equations
of motion for $A_t$ and $A_\rho$ are %
\be %
 (2 \pi \alpha')
 F_{t\rho} = \frac{g \sqrt{1+{\sigma'}^2}}{\sqrt{g^2 + (2 \pi
 \alpha')^2 \rho^{6} (1+(2 \pi \alpha')^2 H B^2)}}.
\ee %
Notice that the above assumption for the currents satisfies current
conservation equation \cite{paper}. Moreover the charge associated with this current is %
\be\label{charge} %
 Q=\int d^7\xi \sqrt{-g} ~j^t,
\ee %
where $Q$ is the number of quarks or equivalently open string endpoints
on the D7-brane. Hence the baryon number density can be defined as %
\be %
 n_B=\frac{Q}{{\rm{V}}_3 N_c},
\ee %
and using \eqref{charge} we finally have \cite{paper} %
\be\label{source}
 g(x^-) = (2/\pi)(2\pi\alpha')^4 \lambda n_B(x^-).
\ee %

Mesons in the AdS/CFT duality are realized as small fluctuations of
the shape of the flavour D7-brane in the transverse directions
\cite{Kruczenski}. In order to find the action describing the
dynamics of scalar mesons we assume the transverse directions have the form
\be %
 x^I = x_0^I + y^I,
\ee %
where $x_0^I$ are the classical solutions
denoting the shape of the D7-brane and $y^I$ are the small
fluctuations. In the presence of the magnetic field, $B$, and the gauge field strength, $F_{t\rho}$, the
action for mesons is obtained by expanding the DBI action for small
fluctuations. Up to second order in $y^I$ the action becomes
\be\begin{split} %
S_{DBI}&= S_0 + S_1 + ... \cr &=S_0 - \frac{\mu_7}{2} \int d^{8}\xi
\sqrt{\gamma_0}~\bigg{(} \gamma_0^{ab} M_{ba} + \gamma_0^{ab}
N_{ba}\cr &\hspace{4.5 cm}-\frac{1}{2} \gamma_0^{ab}
M_{bc}\gamma_0^{cd} M_{da}\ + \frac{1}{4} (\gamma_0^{ab} M_{ba})^2 +
...\bigg{)},
\end{split}\ee
where
\bse\begin{align}
\gamma_{0~ab} &= G_{ab} + G_{IJ} \partial_a x_0^I \partial_b x_0^J + (2 \pi \alpha') F_{ab}, \\
M_{ab} &= \partial_I G_{ab} y^I + G_{IJ} \partial_a x_0^I \partial_b y^J + G_{IJ} \partial_a y^I \partial_b x_0^J
+ \partial_K G_{IJ} \partial_a x_0^I \partial_b x_0^J y^K, \\
N_{ab} &= \frac{1}{2} \partial_I \partial_J G_{ab} y^I y^J
+ G_{IJ} \partial_a y^I \partial_b y^J + \partial_K G_{IJ} \partial_a x_0^I \partial_b y^J y^K\nonumber \\
&+\partial_K G_{IJ} \partial_a y^I \partial_b x_0^J y^K + \frac{1}{2} \partial_K \partial_L G_{IJ} \partial_a x_0^I \partial_b x_0^J y^K y^L,
\end{align}\ese
and
\be %
S_0 = - \mu_7 \int d^{8}\xi \sqrt{\gamma_0}.
\ee %
One can easily observe that if $B$ is set to zero in this set-up which means that the classical solution $x_0^I$ is constant and not coordinate-dependent, the action in \cite{Hashimoto,paper} is recovered. It should be mentioned that the first order terms in $y^I$ are cancelled due to their equation of motion. Note also that the CS action does not contribute since the gauge field and the four-form field have common indices.

In our case we consider the profile of the transverse directions as
\be%
\sigma=\sigma_0(\rho) + \chi(\xi),
\ee%
and $\varphi=0$. We choose $\s_0(\rho)$ to be a
massless solution of the equation of motion \eqref{EOM} and $\chi$
is a small fluctuation around this solution. After some long
calculations the part of the action which describes the kinetic term
for the fluctuations in $\chi$ reduces to
 \be\begin{split}%
 S_1 &=
-\frac{\mu_7}{2} \int d^8\xi~ G_{\s\s} \Big[\big(1+\A^2 H
B^2\big)\big(1-\A^2 F_{t\rho}^2 + \s_0'^2\big)\Big]^{-\frac{1}{2}}\cr
&\hspace{4 cm}\times\sqrt{- \det {\hat s_0}}~ {\hat s_0}^{ab}
~\partial_a \chi \partial_b \chi,
\end{split}\ee%
where we have defined
\be%
{\hat s_0}^{ab} = \frac{1-\A^2 F_{t\rho}^2}{1-\A^2 F_{t\rho}^2 + \s_0'^2}~ s_0^{ab} - \theta_0^{a\rho} \theta_0^{b\rho} \s_0'^2 G_{\s\s}.
\ee%
$s_0^{ab}$ and $\theta_0^{ab}$ are the symmetric and anti-symmetric parts of $\gamma_0^{ab}$, respectively
\be%
\gamma_0^{ab} = s_0^{ab} + \theta_0^{ab}.
\ee%
In our set-up the only nonzero component for $\theta_0^{ab}$ is
\be%
\theta_0^{t\rho} = \frac{\A F_{t\rho}}{1-\A^2 F_{t\rho}^2 + \s_0'^2}.
\ee%
Hence the action $S_1$ for the scalar mesons can be written more neatly as
\be%
S_1= \frac{-1}{2} \int d^8 \xi \sqrt{-det {\tilde s}} ~{\tilde s}^{ab} ~\partial_a \chi \partial_b \chi,
\ee%
where the explicit form of the metric components become
\be\begin{split} %
 -{\tilde s}_{tt} &= \mu_7^{\frac{1}{3}} ~H^{\frac{-1}{3}} ~\big(1+\A^2 H B^2\big)^{\frac{-1}{6}}~ \big(1-\A^2 F_{t\rho}^2 + \s_0'^2\big)^{\frac{5}{6}},\cr
 {\tilde s}_{\rho\rho} &=  \mu_7^{\frac{1}{3}} ~H^{\frac{2}{3}} ~\big(1+\A^2 H B^2\big)^{\frac{-1}{6}}~ \big(1-\A^2 F_{t\rho}^2 + \s_0'^2\big)^{\frac{5}{6}},\cr
 {\tilde s}_{11} &= {\tilde s}_{22}=  \mu_7^{\frac{1}{3}} ~H^{\frac{-1}{3}} ~\big(1+\A^2 H B^2\big)^{\frac{5}{6}}~ \big(1-\A^2
 F_{t\rho}^2 + \s_0'^2\big)^{\frac{-1}{6}},\cr
 {\tilde s}_{33} &=  \mu_7^{\frac{1}{3}} ~H^{\frac{-1}{3}} ~\big(1+\A^2 H B^2\big)^{\frac{-1}{6}}~ \big(1-\A^2 F_{t\rho}^2 + \s_0'^2\big)^{\frac{-1}{6}},\cr
 {\tilde s}_{\alpha\beta} &=  \mu_7^{\frac{1}{3}} ~H^{\frac{2}{3}}~\rho^2 ~\big(1+\A^2 H B^2\big)^{\frac{-1}{6}}~ \big(1-\A^2 F_{t\rho}^2 +
 \s_0'^2\big)^{\frac{-1}{6}} ~{\cal{G}}_{\alpha\beta}.
\end{split}\ee %
$\theta^\alpha(\alpha = 1,2,3$) are the angular variables on $S^3$
and ${\cal{G}}_{\alpha\beta}$ is the metric on the unit
3-sphere. The surface of the presumably apparent horizon, which is defined
locally as a surface whose area variation vanishes along the null ray which is
normal to the surface, is given by%
\be\begin{split}\label{app}
V &=\int d^3x d^3\theta \left(\prod_{i=1}^{3}\tilde{s}_{ii}\prod_{\alpha=1}^{3}\tilde{s}_{\alpha\alpha}\right) \cr
&={\rm{V}}_3 {\rm{Vol}}(\Omega_3)
~\frac{\mu_7 H^{\frac{1}{2}}}{\A}
~~\sqrt{\frac{g^2 + \A^2 \rho^6
(1+\A^2 H B^2)}{(1+\s_0'^2)}}~,
\end{split}\ee %
where this should satisfy
\be %
 dV|_{dt=-dz}=0.
\ee %
In order to simplify the calculations we
obtain the equation for the apparent horizon using the derivative of the square of
(\ref{app}) and get %
\be\begin{split}\label{master} 2
&(1+\s_0'^2)^{\frac{1}{2}} g {\dot g} - g^2 \bigg{(}\frac{2}{R^2}
(\s_0\s_0'+\rho) +H^{\frac{-1}{2}} \frac{\s_0'\s_0''}{1+\s_0'^2}\bigg{)} \cr
&-\A^2 \rho^6 \bigg{(} \frac{2}{R^2} (\s_0\s_0'+\rho) + H^{\frac{-1}{2}}
\frac{\s_0'\s_0''}{1+\s_0'^2} -3 H^{\frac{-1}{2}} \rho^{-1}\bigg{)} \cr
&+\A^4 \rho^6 H B^2 \bigg{(} \frac{-4}{R^2} (\s_0\s_0'+\rho) -
H^{\frac{-1}{2}} \frac{\s_0'\s_0''}{1+\s_0'^2} +3 H^{\frac{-1}{2}}
\rho^{-1}\bigg{)}=0,
\end{split}\ee
where ${\dot g} = \partial_{x^-} g$. We will use this equation in the following subsections to obtain the thermalization time observed on the boundary after we set the functional form of $g(x^-)$. The calculations have been done for the small and general values of the magnetic field, separately.

\subsection{Weak Magnetic Field Limit}\label{hajar}
In this subsection we are going to study the meson thermalization
time for the small values of constant external magnetic
field, \textit{i.e.} $2\pi\alpha' B \ll 1$. As it was mentioned in
section \ref{hajar1} one can set $\s_0(\rho)$ approximately to zero for weak magnetic field. This will simplify the calculations significantly.  The equation
\eqref{master}  reduces to
\be %
 4R^2\rho^{-1}g\dot g-4g^2+2(2\pi\alpha')^2\rho^6-2(2\pi\alpha')^4\rho^2 R^4 B^2=0,
\ee %
or equivalently
\be\label{master1} %
 \frac{\lambda-2\pi^2z^4B^2}{16 n_B^2 \pi^4 z^6}-4 y^2+4 z y \dot{y}=0,
\ee %
where $g(x^-)=g_{max}~y(x^-)$ and $g_{max}$ was introduced in
\eqref{source}. In order to determine the thermalization time we have to specify the form of the function $y(x^-)$. Introducing different
appropriate functions for $y(x^-)$ lead to various forms of baryon
injection into the system. This function increases form zero to a
constant $g_{max}$ during baryon injection and remains constant
after that. A suitable function is \cite{Hashimoto,paper} %
\bea \label{g}
 g(x^-)=\left\{%
\begin{array}{ll}
    0, & x^-<0 \\
    g_{max}\ \omega x^-, & 0<x^-<\frac{1}{\omega} \\
    g_{max}, & \frac{1}{\omega}<x^- \\
\end{array}%
\right.
\eea %
where $\omega x^-$ is a dimensionless combination. The time-dependent injection of baryons leads to the horizon formation on the probe brane where its location can be obtained from \eqref{master1}. The results
are
\bse\begin{align}\label{timesacle1} %
t&=\frac{3z}{2}\pm\frac{\sqrt{\lambda
    +16n_B^2 \pi^4 z^8\omega^2-2\pi^2 z^4B^2}}{8n_B \pi^2 z^3 \omega}~,\hspace{1.53 cm} t<z+\frac{1}{\omega}~,\\
    \label{timesacle2}z^4&=\frac{ B^4+\nu^{-\frac{1}{3}} B^2(B^6 - 3072 n_B^4 \pi^2 \lambda)
    + \nu^{\frac{1}{3}}}{3072 n_B^4 \pi^{4}}
    ,\hspace{1.48 cm} t>z+\frac{1}{\omega}~,
\end{align} %
\ese %
where
\be\begin{split} %
\nu=B^{12}-& 4608 B^6 n_B^4 \pi^2  \lambda +
  24576\bigg(144 n_B^8 \pi^{4} \lambda^2 \cr
  +&\sqrt{6} n_B^6 \pi^3 \lambda^{3/2}
  \sqrt{3456 n_B^4 \pi^2 \lambda-B^6}\bigg).
\end{split}\ee %
\begin{figure}[ht]
\begin{center}
\includegraphics[width=2.9 in]{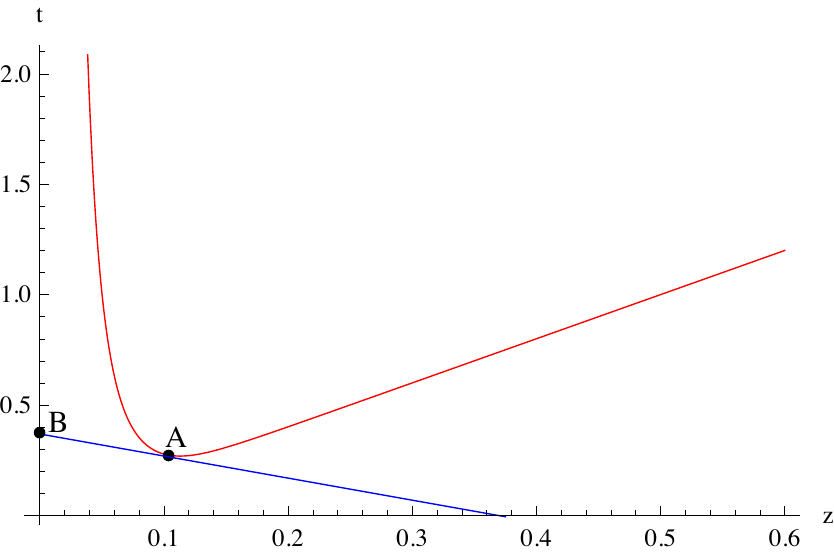}
\caption{ The location of the apparent horizon for $t < z +
1/\omega$ in the $t-z$ plane for $(2\pi\alpha')B=0.09$. The red
curve represents \eqref{timesacle1} and the blue line shows the
earliest tangent null ray.\label{t-z1}}
\end{center}
\end{figure}

\begin{figure}[ht]
\begin{center}
\includegraphics[width=2.7 in]{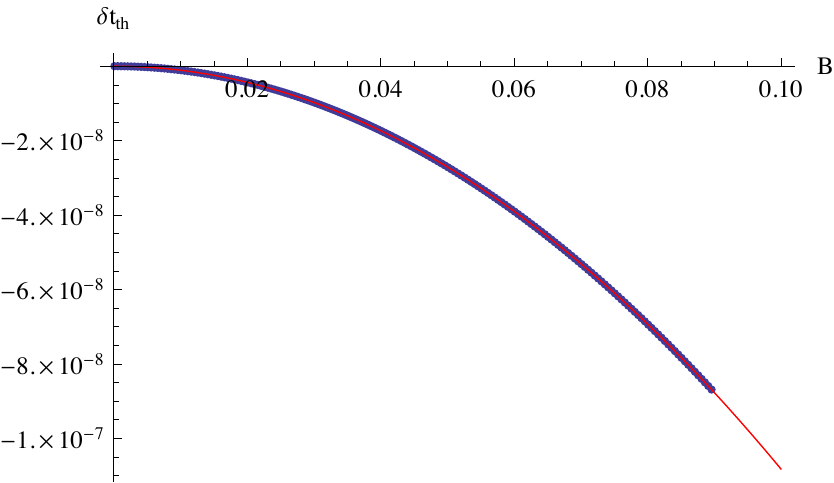}
\hspace{.2 cm}
\caption{$\delta t_{th}$ for $t < z+\frac{1}{\omega}$ is plotted in terms of $B$ in the weak magnetic field limit. The red curve shows the fitted function.\label{hajar11}
}
\end{center}
\end{figure}

In figure \ref{t-z1} the red curve represents the location of the apparent
horizon for $t<z+1/\omega$, \eqref{timesacle1}\footnote{In
order to plot this figure we have set the parameters as
\cite{Hashimoto,paper}: $n_B=34$, $\lambda=10$, $\omega=10$ and $R=1$. We will use these numbers throughout the paper to plot
different figures.}. The thermalization time observed on the boundary is defined as the time where the earliest null ray tangent to the curve reaches the boundary. This has been shown by the point B in figure \ref{t-z1} where the null ray (blue line) is tangent to the curve at point A. Hence in the light cone coordinates the
thermalization time is %
\be
 t_{th}^{B=0.09}=t_A+z_A.
\ee %
It is evident from the above relations that the value of the thermalization time depends on
the external magnetic field. In figure \ref{hajar11}(left) the
difference between the thermalization time in zero and non-zero magnetic field (the blue curve), $\delta
t_{th}=t_{th}^B-t_{th}^{B=0}$, is numerically plotted as a function of
the external magnetic field. It is clear that $\delta
t_{th}$ becomes more and more negative as the magnitude of the magnetic
field increases. In other words the thermalization of
mesons or meson dissociation happens more rapidly in the presence of the magnetic field. Interestingly the magnetic field dependence of the thermalization time is precisely fitted with %
\be\label{fit1}
 t_{th}^B=\bigg(1-29\times 10^{-6} (2\pi\alpha')^2B^2\bigg)t_{th}^{B=0},
\ee %
where $t_{th}^{B=0} \sim \big(\frac{\lambda}{n_B^2\omega^2}\big)^{1/8}$ \cite{Hashimoto,paper}.
The equation \eqref{fit1} reveals that $\delta t_{th}$ is proportional to $\alpha'^2
B^2$ (the red curve in figure \ref{hajar11}).

\begin{figure}[ht]
\begin{center}
\includegraphics[width=2.5 in]{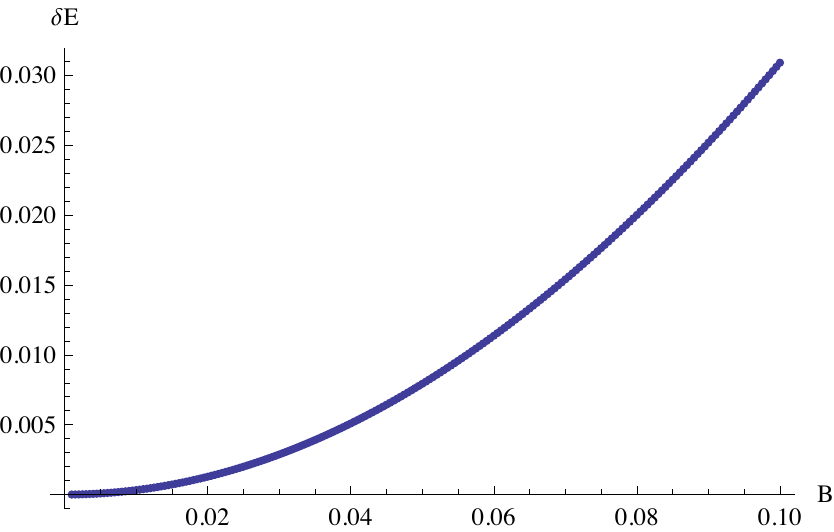}
\caption{The energy of the massless embedding in terms of $B$ in the weak magnetic
field limit. \label{hajar12}}
\end{center}
\end{figure}

The decrease in the thermalization time may be explained by comparing the initial energy of the system (before the baryon injection) with and without the magnetic field.
For
a static solution the energy
is equal to the negative of the action evaluated in that solution. Therefore using
\eqref{hajaraction} the energy density,
$E=\frac{-S_{DBI}}{\mu_7 {\rm{V}}_3{\rm{Vol}}(\Omega_3)\int dt}~,$
for a configuration with nonzero magnetic field is
\be\label{hajarB} %
 E=\int_0^\infty d\rho~\rho^3\sqrt{1+\frac{(2\pi\alpha')^2 B^2}{\rho^4}},
\ee %
and hence %
\be %
\delta E=E-E_0,
\ee %
where $E_0$ is the energy density when $B=0$. The equation \eqref{hajarB} shows that the magnetic field increases the value
of the energy density. This fact is numerically shown in
figure \ref{hajar12}. Therefore a solution with a larger value
of the magnetic field has a larger value of initial energy density and by injecting the energy into the system it would be easier for mesons to dissociate.  Hence
$t_{th}^{B_1}<t_{th}^{B_2}$ for $B_1>B_2$.

\begin{figure}[ht]
\begin{center}
\includegraphics[width=1.7 in]{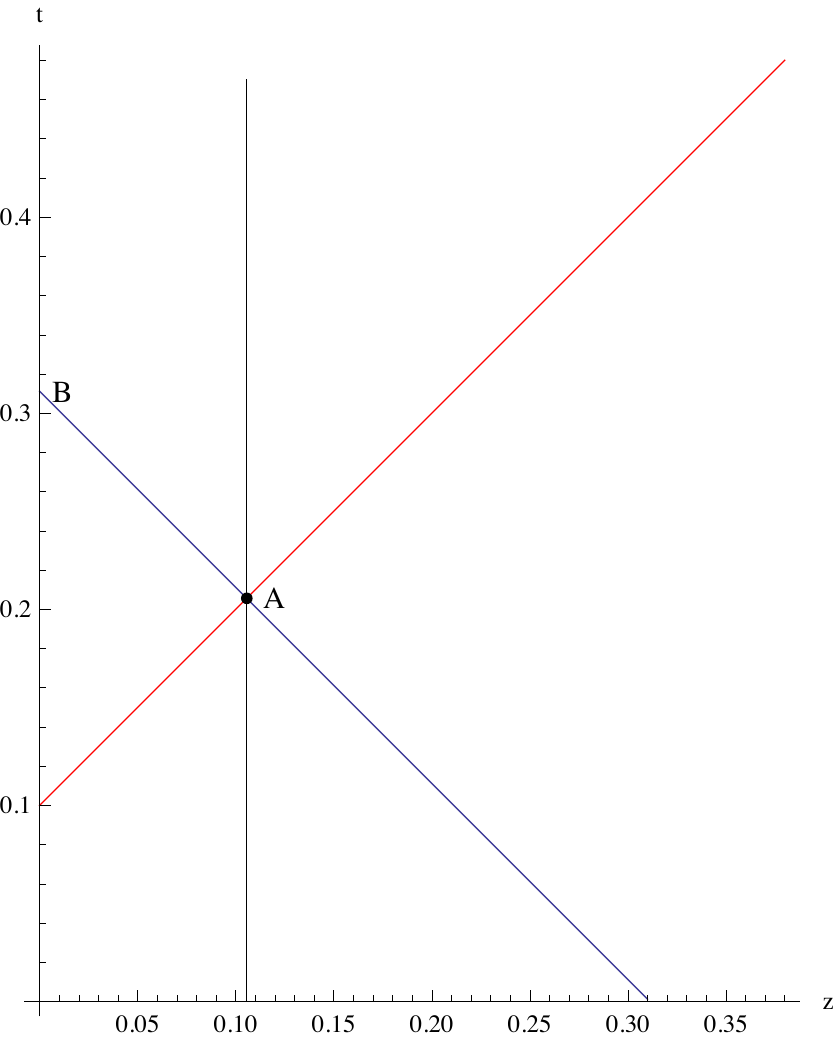}
\hspace{.8 cm}
\includegraphics[width=2.7 in]{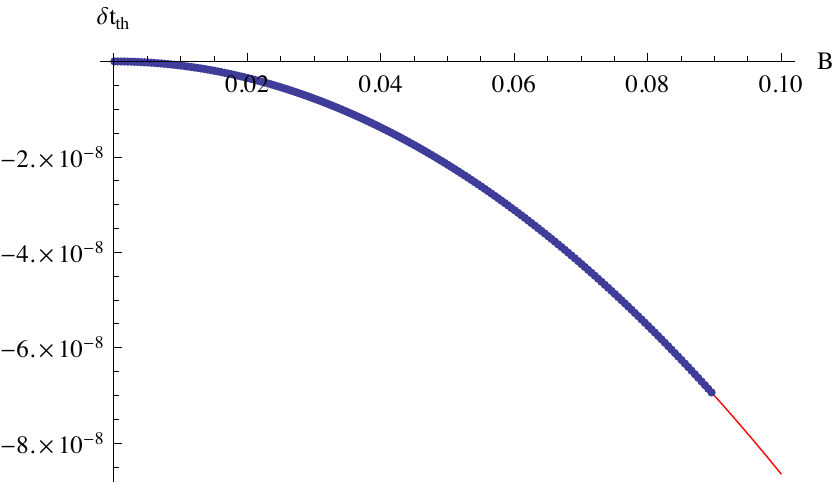}
\caption{ Left: The location of the apparent horizon for $t >z +
1/\omega$ in the $t-z$ plane for $(2\pi\alpha') B=0.09$. The black, red and blue lines represent \eqref{timesacle2}, $t=z+\frac{1}{\omega}$ and the earliest null ray which reaches the boundary. Right:  $\delta t_{th}$ for $t > z+\frac{1}{\omega}$ is plotted in terms of $B$ in the weak magnetic field limit. The red curve shows the fitted function.\label{t-z2}}
\end{center}
\end{figure}

Depending on different values of the parameters, the thermalization
may occur after the baryon injection has been ceased \textit{i.e.}
$t>z+1/w$. Then the equation describing the location of the apparent
horizon on the flavour D7-brane is given by %
\be\label{hajar-master}
 \frac{\lambda-2\pi^2z^4B^2}{16 n_B^2 \pi^4 z^6 }-4 y^2=0.
\ee %
The solution to this equation has been previously found in
\eqref{timesacle2}. In figure \ref{t-z2}(left), $t=z+1/\omega$ and
\eqref{timesacle2} are shown by red and black lines, respectively.
Similar to what we discussed before the point B represents the
earliest time that a boundary observer sees the apparent horizon
formation on the probe brane (point A). The blue line shows an
outgoing null ray form the point A
to the boundary observer. Therefore the thermalization time is %
\be\begin{split}
 t_{th}^{B=0.09}=t_A+z_A,  \cr
 t_A=z_A+\frac{1}{\omega},
\end{split}\ee %
where the value of $z_A$ is given by \eqref{timesacle2}. The same analysis done before to see how the thermalization time changes with magnetic field can be applied here.
 In figure \ref{t-z2}(right) the thermalization time has been plotted with respect to the magnetic field and can be fitted by
\be %
\label{fitweak2}
 t_{th}^B=\bigg(1-28\times 10^{-6} (2\pi\alpha')^2B^2\bigg)t_{th}^{B=0},
\ee %
where $t_{th}^{B=0}=\big(\frac{\lambda}{n_B^2}\big)^{1/6}$ \cite{Hashimoto,paper}. Interestingly we observe that although $t_{th}^{B=0}$ is different in \eqref{fitweak2} and \eqref{fit1}  the dependence of the thermalization time on $B$ is the same for both cases. It looks like, since they both appear in the sane regime of small $B$, only the time-scale of thermalization is different.

\begin{figure}[ht]
\begin{center}
\includegraphics[width=2.9 in]{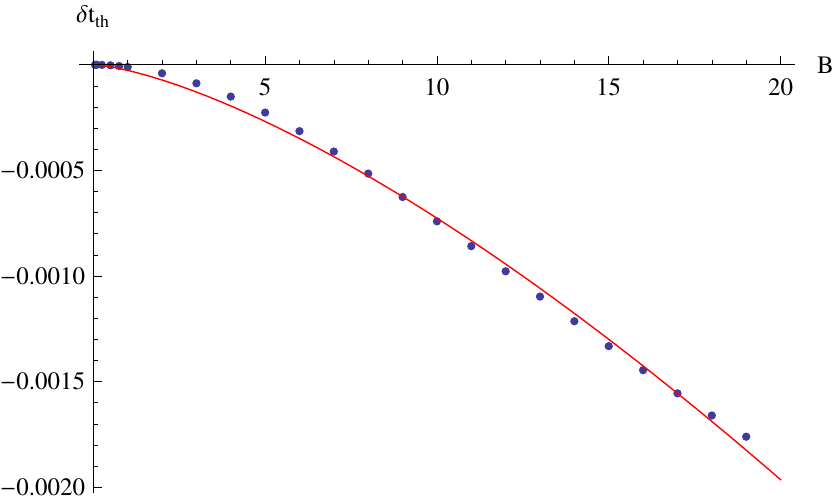}
\caption{ $\delta t_{th}$ for $t < z+\frac{1}{\omega}$ is plotted in terms of $B$ for general values of $B$. The red curve shows the fitted function.
 \label{general}}
\end{center}
\end{figure}

\subsection{General Magnetic Field}
In the previous section we investigated weak magnetic field limit in which we
could assume $\sigma_0(\rho) \approx 0$. Now we would like to generalize it to arbitrary values of $B$
where we have to use \eqref{master} and try to solve it numerically. We follow exactly the same
steps as the previous section and the result for the thermalization time with respect to the value of
the magnetic field has been shown in figure \ref{general}. The blue dots show our numerical results.
It can be fitted by (red curve in figure \ref{general})
\be\label{fit2}
 t_{th}^B=\bigg(1-4\times 10^{-5} (2\pi\alpha')^2B^{1.435}\bigg)t_{th}^{B=0}.
\ee %
Comparing this with \eqref{fit1} and \eqref{fitweak2} we observe that as we increase the magnitude of the
magnetic field the power of $B$ in the fitted equation deviates from the power $2$ of $B$ in those equations. Note that similar to the weak magnetic
field limit the thermalization time decreases as one increases the magnetic field. Such an observation has been also reported in non-commutative SYM \cite{Edalati:2012jj}.

\section*{Concluding Remarks}
\begin{itemize}
\item The main result is that the meson thermalization(dissociation) happens more rapidly in the presence of external magnetic field. The result shown in figure \ref{hajar12} might be used to explain this observation. It indicates
that the initial energy of the system increases as we increase the magnetic field on the probe brane. The figure includes only small values of the magnetic
field but this argument can be generalized to more general values of it. Therefore, due to the increase of the initial energy, we can conclude that as we inject the energy to the system by the baryon
injection the mesons can dissociate more easily.

However we have seen that the above observation is true up to a specific value of the
magnetic field which seems to be around $20$ and then this behaviour reverses. It seems that very large values of the magnetic field distorts
the system in a way that the above argument can not be applied. In fact the theory is strongly coupled and the underlying dynamics remains unclear. Similar
observation has been done in superconductors where there is a critical value of the magnetic field after which the superconducting phase doesn't
happen. Also the phase transition to the superconducting phase happens faster as the magnetic field is raised \cite{Nakano:2008xc}.

\item In weak magnetic field limit $\delta t_{th}\propto B^2$ and for general case  $\delta t_{th}\propto B^{1.435}$.
Therefore we can see the inclusion of larger values of the external magnetic field on the probe brane will lower the power of $B$ in the relation
for $\delta t_{th}$.

\end{itemize}


\section*{Acknowledgement}
We would like to thank D. Allahbakhshi, S. Das, A. E. Mosaffa, A. O'Bannon, M. M. Sheikh-Jabbari and U. A. Wiedemann for useful discussions. We also thank M. Alishahiha for fruitful comments. We are grateful to J. Barandes for some mathematica tips. Both authors would like to thank CERN theory division for their hospitality during their visits.

\end{document}